\documentclass[amsmath,amssymb,showpacs,draft,preprint,floatfix]{revtex4}
\usepackage{graphicx}
\usepackage{latexsym}
\usepackage{amsmath}
\usepackage{amsfonts}
\usepackage{amssymb}
\begin{document}


\title{Entropy operator and  associated Wigner function}
\author{ H. Moya-Cessa}
\affiliation{ INAOE, Coordinaci\'on de Optica, Apdo. Postal 51 y
216, 72000 Puebla, Pue., Mexico}

\begin{abstract}
We show how the entropy operators for two subsystems may be
calculated. In the case of the atom-field interaction we obtain
the associated Wigner function for the entropy operator  for the
quantized field.
\end{abstract}

\maketitle

\section{Introduction}
It is well known that when two subsystems $A$ and $B$ interact the
entropy of the total system and the subsystem entropies obey a
triangle inequality\cite{Arak}
\begin{equation}
|S_A-S_B|\le S_{AB} \le S_A+S_B.
\end{equation}%
This result implies that if both subsystems are initially in pure
states, the total entropy is zero and both subsystems entropies
will be equal after they interact. Here we would like to arrive to
the result that, if initially the two subsystems are in pure
states {\it any} function the density matrix of  subsystem $A$ is
equal to the function of the density matrix of subsystem $B$, and
in particular obtain that both subsystems entropies are equal,
without using the Araki-Lieb theorem. We would also like to find
the entropy operator
\begin{equation}
\hat{S}_{A(B)}=-\ln\hat{\rho}_{A(B)},
\end{equation}
for any of the subsystems. In particular we will consider later a
two-level atom-field interaction, but the results may be
generalized to other kind of subsystems, for instance atom-atom
interaction, atom-many atoms interaction, $N$-level atom-field
interaction, etc.

The density matrix for a two-level system interacting with
another subsystem $B$ is given by
\begin{equation}
\hat{\rho}=\left(
\begin{array}{ll}
|c\rangle\langle c| & |c\rangle\langle s| \\
|s\rangle\langle c| & |s\rangle\langle s|
\end{array}
\right) \label{density-m}
\end{equation}
where $|c\rangle$ ($|s\rangle$) is the unnormalized wave function
of the second system corresponding to the excited (ground) state
of the two level system. From the total density matrix we may
obtain the subsystem density matrices as

\begin{equation}
\hat{\rho}_A=\left(
\begin{array}{ll}
\langle c|c\rangle & \langle s|c\rangle \\
\langle c|s\rangle & \langle s|s\rangle
\end{array}
\right)\equiv \left(
\begin{array}{ll}
\rho_{11} & \rho_{12} \\
\rho_{21} & \rho_{22}
\end{array}
\right),
\end{equation}
and
\begin{equation}
\hat{\rho}_B=|c\rangle\langle c| +|s\rangle\langle s| .
\label{density-f}
\end{equation}
From the  density matrix for subsystem $B$ one can not see a clear
way to calculate the entropy operator as powers of $\rho_B$ get
complicated to be obtained. To make the calculation easier  we
state that {\it if two subsystems are initially in pure states,
after interaction, the trace of any function of one of the
subsystems density matrix  is equal to the trace of the function
of the other subsystem's density matrix.}

We can prove this, by using the following relation valid for two
interacting systems that before interaction were in pure states
\begin{equation}
\hat{\rho}_B^{n+1}=Tr_A\{\hat{\rho}(t)\hat{\rho}_A^n(t)\},
\label{relation}
\end{equation}
with this relation it is easy to show that
\begin{equation}
Tr_B\{\hat{\rho}_B^{n+1}(t)\}=Tr_A\{\rho_A^{n+1}(t)\}.
\end{equation}
 In particular with the
expression (\ref{relation}) we  demonstrate that $S_B=S_A$. Of
course it is also true that
\begin{equation}
\hat{\rho}_A^{n+1}=Tr_B\{\hat{\rho}(t)\hat{\rho}_B^n(t)\}.
\label{relation2}
\end{equation}
\section{Proving $\hat{\rho}_B^{n+1}=Tr_A\{\hat{\rho}(t)\hat{\rho}_A^n(t)\}$ by induction}
We can prove relation (\ref{relation}) by induction. For this we
need to find $\hat{\rho}_A^{n}$, to this end we write
\begin{equation}
\hat{\rho}_A=\frac{{\bf 1}}{2}+ \left(
\begin{array}{ll}
\frac{\delta}{2} & \rho_{12} \\
\rho_{21} & -\frac{\delta}{2}
\end{array}
\right)\equiv \frac{{\bf 1}}{2}+\hat{R},
\end{equation}
where $\delta=\rho_{11}-\rho_{22}$ and ${\bf 1}$ is the $2\times
2$ unit density matrix. We can then find simply
\begin{equation}
\hat{\rho}_A^n=\left(\frac{{\bf 1}}{2}+\hat{R}\right)^n=
\sum_{m=0}^n \left(
\begin{array}{l}
n \\ m
\end{array}
\right) \frac{1}{2^{n-m}}\hat{R}^m .
\end{equation}

We split the above sum into two sums, one with odd powers of
$\hat{R}$ and one with even powers. We also use that

\begin{equation}
\hat{R}^{2m}=\epsilon^{2m}, \qquad
\hat{R}^{2m+1}=\frac{\hat{R}}{\epsilon}\epsilon^{2m+1}
\end{equation}
with
$\epsilon=\left(\frac{\delta^2}{4}+|\rho_{12}|^2\right)^{1/2}$.
Therefore we can write
\begin{equation}
\hat{\rho}_A^n=\frac{1}{2}
\left[\left(\frac{1}{2}+\epsilon\right)^n+\left(
\frac{1}{2}-\epsilon\right)^n \right]{\bf 1}+
\frac{\hat{R}}{2\epsilon}\left[\left(\frac{1}{2}+\epsilon\right)^n-\left(
\frac{1}{2}-\epsilon\right)^n \right]
\end{equation}
In terms of $\rho_A$ the above equation is written as
\begin{eqnarray}
 \hat{\rho}_A^n= G(n)\hat{\rho}_A
- || \hat{\rho}_A || G(n-1) {\bf 1}\label{tothen}
\end{eqnarray}
 where
\begin{eqnarray}
G(n)=\frac{1}{2\epsilon}\left[\left(\frac{1}{2}+\epsilon\right)^n-\left(
\frac{1}{2}-\epsilon\right)^n\right]
\end{eqnarray}
 with the determinant
$||\hat{\rho}_A(t)||=\frac{1}{4}-\epsilon^2$.
 Note that we have written  $\rho_A^n$ in terms of $\rho_A$
and the unity matrix. We could have done it via Cayley-Hamilton
theorem\cite{Alle}, which states that any (square) matrix obeys
its characteristic equation, i.e. a $2\times 2 $ matrix may be
written as we did, as the square of the matrix (and any other
power of it) may be related to the matrix and the unity matrix.

Then  relation (\ref{relation}) my be proved by induction:  for
$n=1$, we have
\begin{equation}
\hat{\rho}_B^2=Tr_A\{\hat{\rho}(t)\hat{\rho}_A(t)\}.
\end{equation}
We now assume  it to be correct for $n=k+1$
\begin{equation}
\hat{\rho}_B^{k+1}=Tr_A\{\hat{\rho}(t)\hat{\rho}^k_A(t)\}
\end{equation}
and prove (\ref{relation}) for $n=k+1$. By using (\ref{tothen}) we
can write
\begin{eqnarray}
 \hat{\rho}_B^{k+1}=
 \hat{\rho}_B^2G(k)-\hat{\rho}_B||\hat{\rho}_A(t)||G(k-1)
\end{eqnarray}
Note that any power of $\hat{\rho}_B$ may be written in terms of
$\hat{\rho}_B$ and $\hat{\rho}_B^2$ ($k\ge2$). Multiplying the
above equation by $\hat{\rho}_B$ we obtain
\begin{eqnarray}
 \hat{\rho}_B^{k+2}=
 \hat{\rho}_B^3G(k)-\hat{\rho}_B^2||\hat{\rho}_A(t)||G(k-1)
 \label{18}
\end{eqnarray}
We obtain $\hat{\rho}_B^3$ from (\ref{relation}) as
\begin{equation}
\hat{\rho}_B^3=Tr_A\{\hat{\rho}(t)\hat{\rho}_A^2(t)\}=Tr_A\{\hat{\rho}(t)[\hat{\rho}_A(t)+(\epsilon^2-\frac{1}{4})]\}
\label{19}
\end{equation}
where for the second equality we have used the Cayley-Hamilton
theorem for the atomic density matrix: $\hat{\rho}_A^2(t)=
\hat{\rho}_A(t)-||\hat{\rho}_A(t)||{\bf 1}$. Inserting (\ref{19})
in (\ref{18}) and after some algebra we find

\begin{eqnarray}
 \hat{\rho}_B^{k+2}=
 \hat{\rho}_B^2G(k+1)-\hat{\rho}_B||\hat{\rho}_A(t)||G(k)
\end{eqnarray}
or
\begin{equation}
\hat{\rho}_B^{k+2}=Tr_A\{\hat{\rho}(t)\hat{\rho}^{k+1}_A(t)\}
\end{equation}
that ends the prove of  relation (\ref{relation}) by induction.
\section{Atomic entropy operator} With the tools we have
developed up to here we can study the two-level atom-field
interaction\cite{Jayn} or equivalently the ion-laser interaction
\cite{Moya3a,Moya3b} and construct atom and field entropy
operators. In the off-resonant atom-field interaction, i.e. the
dispersive interaction, the (unnormalized) states $|c\rangle$ and
$|s\rangle$ read\cite{Haro}
\begin{equation}
|c\rangle=\frac{1}{\sqrt{2}}|\beta e^{-i\chi t}\rangle, \qquad
|s\rangle=\frac{1}{\sqrt{2}} |\beta e^{i\chi t}\rangle
\label{cands}
\end{equation}
where
$|\beta\rangle=e^{-|\beta|^2/2}\sum_{k=0}^{\infty}\frac{\beta^k}{\sqrt{k!}}|k\rangle$
is the initial coherent state for the field, $\chi$ is the
interaction constant. The atom was considered initially in a
superposition of excited and ground
$|\psi_A\rangle=\frac{1}{\sqrt{2}}(|e\rangle+|g\rangle)$.

We write the entropy operator as
\begin{eqnarray}
\hat{S}_A=\ln\hat{\rho}_A^{-1}=\ln(1-\hat{\rho}_A)-\ln
||\hat{\rho}_A||,
\end{eqnarray}
such that the expectation value of $\hat{S}_A$ is the entropy. In
the above expression we have used that
$\hat{\rho}_A^{-1}(t)=\hat{\xi}_A/||\hat{\rho}_A(t)||$ with the
purity operator $\hat{\xi}_A={\bf 1}-\hat{\rho}_A(t)$. We can
develop $\ln(1-x)=-\sum_{n=1}^{\infty}\frac{x^n}{n}$and use
(\ref{tothen}) to find
\begin{eqnarray}
\hat{S}_A=F_1\hat{\rho}_A+F_2 {\bf 1} \label{atent}
\end{eqnarray}
with
\begin{eqnarray}
\nonumber
& &F_1=\frac{1}{2\epsilon}\ln\left(\frac{1-2\epsilon}{1+2\epsilon}\right)\\
& &F_2=-\frac{1}{2}\left[\ln||\hat{\rho}_A(t)||+
\frac{1}{2\epsilon}\ln\left(\frac{1-2\epsilon}{1+2\epsilon}\right)\right]
\end{eqnarray}
 Note that $\hat{S}_A$ is linear in the atomic density matrix
as expected from Cayley-Hamilton's theorem (for $2\times 2$
matrices). From (\ref{atent}) we can calculate the atomic (field)
entropy and the atomic (field) entropy fluctuations
\begin{eqnarray}
\langle\hat{S}_A\rangle=F_2+F_1(1-2||\hat{\rho}_A(t)||),
\end{eqnarray}
and
\begin{eqnarray}
\langle\Delta\hat{S}_A\rangle=\sqrt{\langle\hat{S}^2_A\rangle-\langle\hat{S}_A\rangle^2}=
\ln\left(\frac{1-2\epsilon}{1+2\epsilon}\right)||\hat{\rho}_A(t)||^{1/2}.
\end{eqnarray}

\section{Field entropy operator} We use the expression for
$\hat{\rho}_B^n$ in terms of $\hat{\rho}_A$ to write the field
entropy operator in terms of the atomic density operator
\begin{eqnarray}
\hat{S}_B=Tr_A\{\hat{\rho}(t)\hat{S}_A(t){\hat{\rho}}^{-1}_A(t)\}
\label{ent-field}
\end{eqnarray}
Note that we can write the atomic entropy operator in terms of the
operator used to define concurrence\cite{Woot} because $
\hat{\rho}^{-1}_A(t)=\frac{1}{||\hat{\rho}_A(t)||}\hat{\sigma}_y\hat{\rho}^*_A(t)\hat{\sigma}_y
=\frac{\hat{\tilde{\rho}}_A(t)}{||\hat{\rho}_A(t)||} $ i.e.
\begin{eqnarray}
\hat{S}_B=-\frac{1}{||\hat{\rho}_A(t)||}Tr_A\{\hat{\rho}(t)\hat{S}_A\hat{\tilde{\rho}}_A(t)\}
\end{eqnarray}
By inserting (\ref{atent}) into (\ref{ent-field}) we obtain
\begin{eqnarray}
\hat{S}_B= Tr_A\{\hat{\rho}(t)(F_1 +F_2\hat{\rho}_A^{-1}(t))\}
\end{eqnarray}
using the expression of the inverse of the atomic density operator
in terms of the purity operator,the entropy may be written as
\begin{eqnarray}
\hat{S}_B= Tr_A\{\hat{\rho}(t)(F_1
+\frac{F_2}{||\hat{\rho}_A(t)||}[{\bf 1}-\hat{\rho}_A(t)])\}
\label{29}
\end{eqnarray}
 In terms of the field density matrix the
field entropy operator is
\begin{eqnarray}
\hat{S}_B=\left(F_1+\frac{F_2}{||\hat{\rho}_A(t)||}\right)\hat{\rho}_B(t)
-\frac{F_2}{||\hat{\rho}_A(t)||}\hat{\rho}^2_B(t) \label{S-f}
\end{eqnarray}
from (\ref{density-f}) we can write $\hat{\rho}^2_B(t)$ as
\begin{equation}
\hat{\rho}^2_B=|c\rangle\langle c|\langle c|c\rangle
+|s\rangle\langle s| \langle s|s\rangle +|c\rangle\langle
s|\langle c|s\rangle +|s\rangle\langle c| \langle
s|c\rangle\label{density-f2}.
\end{equation}
From (\ref{cands}) we have that $\langle c|c\rangle=\langle
s|s\rangle=1/2$ and that $\langle c|s\rangle=\langle
s|c\rangle^*=\exp(-|\beta|^2[1-e^{2i\chi t}])/2$.

The entropy operator for the field is written as
\begin{eqnarray}
\hat{S}_B=
\left(F_1+\frac{F_2}{2||\hat{\rho}_A(t)||}\right)(|c\rangle\langle
c| +|s\rangle\langle s|)- \frac{F_2}{||\hat{\rho}_A(t)||}(\langle
s|c\rangle|s\rangle\langle c| + \langle c|s\rangle |c\rangle
\langle s|)
\end{eqnarray}

 We have chosen  the dispersive interaction so that
all the terms forming the field density matrix and its square are
coherent states, such that we can calculate the Wigner function
associated to (\ref{S-f}) in an easy way. We do it by means of the
formula\cite{Wign}
\begin{equation}
W_S(\alpha)=\sum_{n=0}^{\infty}(-1)^n\langle\alpha,n|\hat{S}_B|\alpha,n\rangle
\end{equation}
where $|\alpha,n\rangle$ are the so-called displaced number states
\cite{Disp}. The explicit expression for the above equation is
\begin{eqnarray}
\nonumber
W_S(\alpha)&=&e^{-2\beta^2-2|\alpha|^2+4\beta\alpha_x\cos\chi t}
\left(F_1+\frac{F2}{2||\hat{\rho}_A(t)||}\right)\cosh(4\beta\alpha_y\sin\chi
t)\\
&-&e^{-2\beta^2-2|\alpha|^2+4\beta\alpha_x\cos\chi
t}\frac{F_2}{2||\hat{\rho}_A(t)||}\cos(2\beta[\beta\sin2\chi
t-2\alpha_x\sin\chi t]).
\end{eqnarray}

\section{Conclusions}

We have shown that the entropy operator for the subsystems may be
obtained and in the case one of those subsystems is the quantized
field. The conection with quasiprobability distribution functions
\cite{Roversi1,Roversi2}, in particular the association to the
Wigner function for the entropy has been obtained.


\begin{thebibliography}{9}
\bibitem{Arak} H. Araki  and E. Lieb, Commun. Math. Phys. {\bf 18},
160 (1970).
\bibitem{Alle} R.B.J.T. Allenby  (1995) "Linear Algebra", Modular Mathematics, Edward
Arnold.
\bibitem{Jayn} E.T. Jaynes and F.W. Cummings, Proc. IEEE {\bf 51}, 81 (1963).
\bibitem{Moya3a}  H. Moya-Cessa and P. Tombesi,  Phys. Rev. A {\bf 61}, 025401 (2000).
\bibitem{Moya3b} H. Moya-Cessa, D. Jonathan and P.L. Knight,{   J. of Mod. Optics} {\bf  50}, 265 (2003).
\bibitem{Haro} L.G. Lutterbach and L. Davidovich, Phys. Rev. Lett {\bf 78}, 2547 (1997).
\bibitem{Woot} W.K. Wootters, Phys. Rev. Lett. {\bf 80}, 2245
(1998).

\bibitem{Wign} W.P. Schleich, "Quantum Optics in Phase Space", (Wiley-VCH, 2001).
\bibitem{Disp} F. A. M. de Oliveira, M. S. Kim, P. L. Knight, and V. Buzek
Phys. Rev. A {\bf 41}, 2645 (1990).
\bibitem{Roversi1} H. Moya-Cessa, S.M. Dutra, J.A. Roversi, and A.
Vidiella-Barranco, { J. of Mod. Optics} {\bf 46}, 555 (1999).
\bibitem{Roversi2} H. Moya-Cessa, J.A. Roversi, S.M. Dutra, and A. Vidiella-Barranco,
Phys. Rev. A {\bf 60},  4029 (1999).
\end{thebibliography}
\end{document}